\documentclass[12pt]{article}
\newcommand{\bmat}{\left(\begin{array}}
\newcommand{\emat}{\end{array}\right)}
\def\NPB#1#2#3{Nucl. Phys. B{#1} (19#2) #3}
\def\PLB#1#2#3{Phys. Lett. B{#1} (19#2) #3}

\def\PRD#1#2#3{Phys. Rev. D{#1} (19#2) #3}

\def\yzero{\smash{\hbox{$y\kern-4pt\raise1pt\hbox{${}^\circ$}$}}}

\def\-{\hphantom{-}}

\def\s2{\frac{1}{\sqrt2}}

\def\beq{\begin{equation}}
\def\eeq{\end{equation}}
\def\beqa{\begin{eqnarray}}  
\def\eeqa{\end{eqnarray}}

\def\Tr{{\rm Tr \,}}
\def\diag{{\rm diag \,}}
\def\IF{\relax{\rm I\kern-.18em F}} 
\def\II{\relax{\rm I\kern-.18em I}}
\def\IP{\relax{\rm I\kern-.18em P}}
\def\IC{\relax\hbox{\kern.25em$\inbar\kern-.3em{\rm C}$}}
\def\IR{\relax{\rm I\kern-.18em R}}

\def\Dsl{\,\raise.15ex\hbox{/}\mkern-13.5mu D} 
\def\IZ{Z\kern-.4em  Z}

\begin{document}
\vspace*{-.6in}
\thispagestyle{empty}
\begin{flushright}
FTUAM-99-37;\ IFT-UAM/CSIC-99-44\\
hep-th/9911499
\end{flushright}
\baselineskip = 16pt

\vspace{.5in}
{\Large
\begin{center}
The second string (phenomenology)
revolution
\end{center}}

\vspace{1in}
\begin{center}
Luis E. Ib\'a\~nez\\
\emph{ Departamento de F\'{\i}sica Te\'orica C-XI \\
and \\ Instituto de F\'{\i}sica Te\'orica  C-XVI,\\
Universidad Aut\'onoma de Madrid \\
Cantoblanco, 28049 Madrid, Spain
}
\end{center}
\vspace{1in}

\begin{center}
\textbf{Abstract}
\end{center}
\begin{quotation}
\noindent
In the past three years our views on how the standard
model of particle physics  could be embedded into
string theory have dramatically changed. 
The heterotic string is no longer  the only possibility
for such an embedding and other perturbative 
(or non-perturbative) corners of M-theory, like Type I or
Type II strings seem now possible. It has also been realized 
that the string scale $M_s$ is not necessary close to the 
Planck scale and could be much smaller, of order the
intermediate scale $\sqrt{M_WM_p}$ or even close to
the weak scale. In addition,  semi-realistic
three generation models have recently been constructed 
starting with Type IIB compact orientifolds. I briefly
discuss some of these developements which represent 
a revolution in our understanding of string phenomenology.

\end{quotation}

\vfil
\centerline{\it To appear in the proceedings of  Strings 99}

\newpage

\pagenumbering{arabic}

\section{Introduction}

Breakthroughs in string theory are often named
 {\it string revolutions}.
The first took place around 1984 when the cancellation of
anomalies in Type I string theory was shown by Green and Schwarz 
and the two hererotic strings were constructed \cite{84rev} .
 The second revolution around 1994 when the importance of
weak/strong coupling dualities  \cite{90rev} were generally apreciated,
leading to a unification of all five supersymmetric string theories
with 11-D supergravity \cite{94rev} . If string theory is correct, it should 
contain as a certain (low energy) limit the observed standard
model (SM) of particle physics. String phenomenology 
\cite{rev} is the study
of the possible avenues by which the $SU(3)\times SU(2)\times U(1)$
SM could be embedded in string physics.  The string revolutions
also imply string phenomenology revolutions, which typically take
place a few years later. This is because it takes some time to
construct explicit semi-realistic string vacua with the new 
 techniques. We are now living the second
 string phenomenology revolution. Progress is sometimes realizing 
what we do not know but  thought we knew. This is now the case:
we have realized that we do not even know what is the fundamental
energy scale of string theory!
To see what has changed in the last three years or so, let me remind you
the pre-1995 orthodoxy in string phenomenology \cite{rev} :

i) Only the heterotic strings (mostly $E_8\times E_8$) where 
considered as contendants for the unified theory of all interactions.
Four dimensional string vacua with one unbroken SUSY are considered.

ii) The string scale $M_s$ is tied up to the Planck scale $M_p$ by the
heterotic string expression $M_s=gM_p$. Thus string physics has its realm
at energies of order $10^{17}$ GeV.

iii) The weak scale $M_W$ is generated by some non-perturbative 
effect like gaugino condensation in a hidden sector of the theory.
If $<\lambda \lambda >\propto   M_WM_p^2$ the required SUSY-breaking
soft terms $\propto M_W$ would be produced for the SUSY SM particles.

iv) Gauge coupling constants of the MSSM unify at a scale 
$M_X=2\times 10^{16}$ GeV, only slightly below the string scale.

A few  examples of semi-realistic four-dimensional string models
\cite{rev}  
with three generations of quarks and leptons have been constructed
using different techniques ( $Z_N$ orbifolds, free fermionic constructions,
Calabi-Yau compactifications etc.). This is by itself an important achievement
since these  models represent the first unified models of all interactions
including gravity. There is however a loss of predictivity in the models
because in order to make contact with the SM (or the MSSM) one has to
abandon the string theory techniques and analize at the level of the
effective field theory the flat directions in each model.
Thus possible predictions like quark/lepton masses or proton stability
become dependent on the flat field direction chosen.
And  of course, the questions of vacuum degeneracy and dilaton/moduli
stabilization (not to mention the cosmological constant) remain to be solved.

Let us enumerate now some
 of the important points 
(from the phenomenological point of view) which
have changed in the last few years:

i)  End of the heterotic monopoly : other string theories like Type I or Type II
or techniques like F-theory provide for new classes of $D=4$, $N=1$ string vacua
which can lead to new avenues to embedd the observed physics.

ii) It has been realized that the number of extra dimenssions felt by
gauge fields and gravity fields may be in general different.
Indeed, gauge fields in e.g., Type I theory live on the
world-volume of D-branes \cite{polrev} 
which span a number of dimensions
often smaller than the full ten dimenssions felt by the
gravity fields \cite{witten,lykken,untev,anton} .

iii) As a consequence of ii), the string scale $M_s$ has no theoretical
bound and we only have the phenomenological bound $M_s\geq $ 1 TeV
\cite{lykken,untev,anton} .

Concerning the first point, we are only starting to scratch the 
space of vacua offered by the new techniques.
Indeed, with the new techniques available, Type I and Type IIB
strings are as good as the heterotic models from the point of
view of model building (see in particular ref. \cite{aiq} ) 
. On the other hand, any of the string theories
are just perturbative limits of the underlying unique M-theory.
So the question is whether some corner of the M-theory moduli
space sits sufficiently close to the observed SM physics.
I am going to concentrate here on the 
case of the 
 $D=4$ vacua obtained
from Type I theory (or equivalently, Type IIB $D=4$ orientifolds
\cite{bl,afiv,mas} )
\footnote{See talk by B. Ovrut for an alternative which
embeds the SM in strongly coupled heterotic theory.}.
These are interesting because they are examples of explicit, perturbative
 string vacua in which many of the new features in model building 
(like e.g., the possibility of a reduced string scale, large
gauge groups, multiple pseudoanomalous $U(1)$'s etc.) are already
present.
Although  Type I strings were discovered
 well before the heterotic
strings, 
little effort has been devoted in the past to the construction of
Type I four-dimensional string vacua. Of course, smooth Calabi-Yau
compactifications of Type I are possible consistent solutions but
they have no phenomenological interest and, anyway, present no
obvious adventage over the heterotic $SO(32)$ theory.
On the other hand the concept of
D-brane  \cite{polrev}  has provided us with a new 
understanding of Type I string theory .

\section{Type IIB $D=4$, $N=1$ compact orientifolds}

Let us describe a bit how these orientifolds 
\cite{sagnotti,horava,dab} 
are constructed.
In a $D=4$ Type IIB orientifold, the toroidally compactified theory is
divided out by the joint action \cite{sagnotti,horava,gp} 
of a discrete symmetry group $G_1$,
like $Z_N$  together
with a world sheet parity operation $\Omega$, exchanging left and right
movers. The $\Omega $ action can be accompanied by extra operations thus
leading to a  generic  orientifold group $G_1+ {\Omega} G_2$ with
${\Omega}h {\Omega}
h' \in G_1$ for $h,h' \in G_2$.
We will consider 
here  the cases $G_1=G_2$ and $G_1=Z_N$ and such that $D=4$ $N=1$
theories are obtained, when
the twist $\Omega$ is performed on Type IIB compactified on $T^6/G_1$. 
The
allowed orbifold groups, acting crystalographically on $T^6$ leading to 
$N=1$  unbroken supersymmetry were classified in \cite{dhvw}.
The finite list of twists is $Z_3$, $Z_4$, 
$Z_6$, $Z_6$', $Z_8$, $Z_8$', $Z_{12}$ and $Z_{12}$'.
Here the primed twists correspond to diferent implementations
of discrete rotations of the given example.

Orientifolding the closed Type IIB string introduces a Klein-bottle
unoriented
world-sheet. Amplitudes on such a surface contain tadpole divergences.
In order to eliminate such unphysical divergences Dp-branes must be
generically introduced. In this way, divergences occurring in the
open string sector cancel  up the closed sector ones and  produce a
consistent theory. For $Z_N$, with $N$ odd, only
D9-branes are required. They fill the full space-time and six dimensional
compact space. For $N$ even, D$5_k$-branes, with world-volume filling
space-time and the $k^{th}$ complex plane, may be required. This is so
whenever the orientifold group contains the element $\Omega O_i O_j$, for
$k\neq i,j$. Here $O_i$ ($O_j$) is an order two twist of the $i^{th}$
($j^{th}$) complex plane. 

One denotes (see  ref.\cite{afiv} for conventions ) open string
states  by $|\Psi, ab \rangle $, where $\Psi$ refers
to world-sheet degrees of freedom while the $a,b$ Chan-Paton indices are
associated to the open string endpoints lying on D$p$-branes and D$q$-branes
respectively.
These Chan-Paton labels must be contracted with a hermitian matrix
$\lambda ^{pq} _{ab}$ which parametrize the gauge indices.
   The action of an element of the orientifold group
on Chan-Paton factors is achieved by a unitary matrix $\gamma _{g,p}$ such
that $g: \lambda ^{pg} \rightarrow \gamma _{g,p} \lambda ^{pq}
\gamma^{-1}_{g,q}$.  We denote by $\gamma _{k,p}$ the matrix associated to
the $Z_N$ orbifold twist $\theta ^k $ acting on a Dp-brane.
A  generic matrix  $\gamma _{1,p}$  can be written as
$
\gamma_{1,p}=({\tilde \gamma_{1,p}},{\tilde \gamma _{1,p}}^{*})
$
where  $* $ denotes complex conjugation. ${\tilde \gamma }$ is a
$N_p\times N_p$ diagonal matrix given by
\beq
{\tilde \gamma }_{1,p}   =
\diag  (\cdots,\alpha^{NV_j}I_{n_j^p},\cdots, \alpha^{NV_ P} I_{n_P^p})
\label{gp}
\eeq
with $\alpha = {\rm e}^{2i\pi /N}$.
$V_j=\frac{j}N$ with $j=0,\dots, P$ corresponds to an action ``with
vector structure '' ($\gamma ^N=1$) while $V_j= \frac{2j-1}{2N}$ with
$j=1,\dots,P$ describes an action
 ``without vector structure'
($\gamma_{1,p} ^N=-1$).
The gauge fields living on the world-volume of a D$p$-brane have associated
Chan-Paton factors $\lambda ^p$ corresponding to the gauge group $G_p$ 
with $G_9=SO(2N_9)$ and $G_5=Sp(2N_5)$. In Cartan-Weyl  basis such generators
are organized into charged generators $\lambda_a = E_a$, $a=1,\cdots, {\rm
dim}\, G_p - {\rm rank}\, G_p$, and Cartan algebra generators $\lambda_I =
H_I$, $I=1,\cdots, {\rm rank}\, G_p$, such that
$
[H_I, E_a]=\rho_I^aE_a
$
where the (${\rm rank}\, G_p$)-dimensional vector with components $\rho_I^a$
is the root associated to the generator $E_a$.

The matrices  $\gamma_{1,p}$ and its powers represent the action of the
$Z_N$ group on Chan-Paton factors, and they correspond to elements of a
discrete subgroup of the Abelian group spanned by the Cartan generators.
Hence, we can write \cite{afiv} 
$
\gamma _{1,p}= e^{-2i\pi V^p \cdot H } .
$
Thus, this equation defines a (${\rm rank}\, G_p$)-dimensional vector $V^p$
with  coordinates corresponding  to the $V_j$'s defined in (\ref{gp})
above.
 In such a description the massless states are easily
found \cite{afiv} .
 Let us consider the case in which all 5-branes sit at the
origin.
In the $(pp)$ sector the gauge group is obtained by
selecting the root vectors satisfying
$ \rho^a \cdot V^p= 0 {\rm \, mod \,} {\bf Z} $
while  matter states correspond to charged generators with
$ \rho^a \cdot V^p= v_i {\rm \, mod \,} {\bf Z} $.
Here the $v_i$ are the eigenvalues of the $Z_N$ rotation of
the three complex compact dimensions (see ref.\cite{afiv}).
In the  $(95)$ sector the subset of roots of $G_9\times G_5$ of the form 
$
P _{(95)}= (W_{(9)}; W_{(5)})=
({\underline {\pm 1, 0,  \cdots, 0}};{ \underline {\pm 1, 0, \cdots, 0}}) 
$ must be considered.  Matter states are obtained from
the projection
$
P_{(95)} \cdot V ^{(95)}= (s_jv_j +s_kv_k) {\rm \, mod \,} {\bf Z}
$
with $s_j=s_k=\pm \frac1{2}$, plus (minus) sign corresponding to particles
(antiparticles) and  $V^{95}= (V^9; V^5)$.

Twisted tadpole cancelation turns out to be quite restrictive
in these   models. In particular, it was shown in ref.\cite{afiv} 
that the only tadpole-free $Z_N$ orientifolds are 
$Z_3$, $Z_6$, $Z_6'$, $Z_7$ and $Z_{12}$. Furthermore, for all those models
(except $Z_6$ and $Z_{12}$ \cite{abiu} ) tadpole conditions fix completely the
gauge group and massless spectrum (modulo Wilson lines and/or 
moving of 5-branes). All these models lead to $N=1$, $D=4$ consistent
vacua with a chiral anomaly-free spectrum.

Instead of $\Omega $ one can use other $Z_2$ modings which are still
consistent with  $N=1$ SUSY in $D=4$.
Thus, for example one can use as orientifold projector
$(-1)^{F_L}\Omega O_i $  or
$(-1)^{F_L}\Omega O_iO_jO_k$, $i\not = j\not = k\not = i$.
Here  $F_L$ is the world-sheet left-handed fermion number.
In this case  tadpole cancellation conditions
will require in general the presence in the vacuum of 7-branes and
3-branes respectively. There may be three different types of
7-branes, $7_i$, $i=1,2,3$ depending what complex dimension $X_i$ is
transverse to the 7-brane world-volume.
Thus we see that, depending on the orientifold generators, one
can deal with 3-branes, $5_i$-branes, $7_i$-branes
and 9-branes. Not all types may be present simultaneously if
we want to preserve $N=1$ in $D=4$. For a given $D=4$, $N=1$
vacuum with D-p-branes and D-p$'$-branes one must have
$(p-p')=0,\pm 4$.
The number of each type of p-brane in each case is dictated
by tadpole cancellation constraints. These in turn guarantee
the cancellation of gauge anomalies in the effective  
$D=4$, $N=1$ theory. 
T-dualities relate the different types of p-branes present in each given 
vacuum \cite{polrev}. Consider for simplicity the 6-torus as the product of
three two-tori, $T^6=T^2\times T^2\times T^2$ each with compact radii
$R_i$, $i=1,2,3$. Now, it is well known that
a duality transformation $R_i\rightarrow \alpha '/R_i$ transforms
Neumann boundary conditions on the $X_i$ coordinate into Dirichlet
boundary conditions and vice versa \cite{polrev}. This means that e.g.,
a 9-brane will turn into a $7_i$-brane and vice versa under this 
transformation.
Thus given any configuration with certain distribution
of p-branes in the vacuum, there are a number of equivalent
configurations which are obtained from T-dualities.

Given a p-brane in a background with six compact dimensions, open strings  
ending on that p-brane will only have
Kaluza-Klein (KK) states along the compact dimensions with
Neumann boundary conditions.  On the contrary, it will have
winding states only in those compact directions with
Dirichlet boundary conditions. 
 On the other hand, closed strings can have both
KK and winding modes in all compact dimensions. This turns out to be   
important in order to study the structure of mass scales in the theory.

\section{The string scale and the Planck mass}

Let us study   the relationship
between string, Planck and compactification scales in Type I $D=4$
strings of the type described above.
 We consider the dimensional reduction down to four dimensions
obtained by compactification on an orbifold with an underlying compact
torus of the form $T^2\times T^2\times T^2 $.  The three
tori are taken with volumes $(2\pi R_i)^2$, $i=1,2,3$ respectively.
The relevant piece
of the bosonic action of the $D=4$, $N=1$ effective Lagrangian 
for a generic distribution of D-branes has the form:
\beqa
S_{4} \ &  = & \ - \int {{ dx^{4}}\over {2\pi }}
 \sqrt{-g}\  (\  {{R_1^2R_2^2R_3^2 M_s^8}\over {\lambda_I^2}} \ R \ + \
{{R_1^2R_2^2R_3^2M_s^6}\over {\lambda _I}} \ {1\over 4} F^2_{(9)} \nonumber \\
&  + & \ \sum _{i\not= j\not=k\not= i} {{R_i^2 R_j^2 M_s^4}\over
{\lambda_I}} \ {1\over 4} F^2_{(7_k)} \  +  \ \sum _{j=1}^3
{{R_j^2M_s^2}\over {\lambda _I}} \ {1\over 4}  F^2_{(5_j)} \ +
{1\over {\lambda _I}}\ {1\over 4}  F_{(3)}^2  \ +\ ...\  )\ , 
\label{d04}
\eeqa
where $\lambda _I$ is the $D=10$ Type I dilaton, $M_s=1/\sqrt{\alpha'}$
is the Type I string scale and 
 we have displayed the kinetic terms for gauge bosons of the different
groups which may come from the different p-branes, $p=9, 7_k, 5_j, 3$.
As discussed above, not all the different p-brane sectors
should be present in the vacuum if we want to respect $N=1$ SUSY.
 From the above
equation one obtains for the gravitational coupling $G_N$
\beq
G_N \ =\ {1 \over {M_{Planck}^2}} \ =\  { {\lambda_I^2 M_1^2 M_2^2 M_3^2}
       \over {8 M_s^8  }}
\label{planck}
\eeq
and for the gauge couplings $\alpha_p $ for the different
p-branes :
\beqa
 \alpha _9 \ & = &\   { { \lambda _I M_1^2M_2^2M_3^2 } \over
{2 M_s^6 } }     \ ; \
\alpha _{7_i} \ =\    { { \lambda _I M_j^2M_k^2 } \over
{2 M_s^4} } \ \ , i\not= j \not= k\not= i  \nonumber \\
\alpha _{5_i} \  &  = & \    {  { \lambda_I M_i^2 } \over
{2 M_s^2 } } \ \ ;\ \
\alpha _3 \ =\ { {\lambda _I}\over 2 } \ ,
\label{gaugci}
\eeqa
where $M_i=1/R_i$. From the above formulae we observe that,
unlike what happens in the heterotic case, $M_{Planck}$ and
$M_s$ do not need to be of the same order of magnitude \cite{witten}.
Consider  for example the simple isotropic case in which all
compactification radii are taken equal, $R_i=R=1/M_c$. Then one gets 
\cite{witten,lykken,untev,anton,st,imr}
\beq
M_{Planck}^2 \ =\ {{8M_s^8}\over {\lambda _I^2 M_c^6} } \ ;\
\alpha _p \ = \ { {\lambda _I}\over 2} ( { {M_c }\over {M_s} })^{p-3 } \ ,\
p=9,7,5,3
\label{isotr}
\eeq
that combined give the following relationship
\beq
{ {M_c^{(p-6)}} \over {M_s^{(p-7)} } } \ =\ { {\alpha _p M_{Planck} }\over
{\sqrt {2} }  }\  ,
\label{const}
\eeq
 Notice that in principle these equations give us a certain
freedom to play with the values of the Type I string scale $M_s$ and the
compactification scale $M_c$. This is to be compared to the analogous
equation in the perturbative heterotic case where the relation
$M_{string}=\sqrt{\frac{\alpha_X}{8}} M_{Planck}$ fixes the value of the
string scale independently of the compactification scale.

Consider for example the case of a set of 3-branes
in an isotropic compactification. One then has:
$
M_s^4 \ =\ {{\alpha_3}\over {\sqrt{2}}}M_c^3 M_p
$
Thus one can e.g. lower the string scale down to e.g. 1 TeV (
which is the lower phenomenological bound) by chosing $M_c=$ 10 MeV
\cite{untev,anton} .
Notice that the compactification scale $M_c$
may be this low because the charged fields living on the
3-branes have no KK excitations and hence there is no charged 
threshold at the $M_c$ scale.  
Thus the lesson that we learn from these  considerations is 
that we do not really know what the string scale is!
This is quite surprising because one of the most  widely
spread dogmas  about string theory is that its natural
scale is the Planck scale
\footnote{Recently an interesting alternative has been suggested
\cite{rs} in which gravitons are trapped on domain walls 
and the extra dimensions are non-compact.}.

So, what is the string scale?
There are a number of natural options for the string scale $M_s$:

{\bf i) $M_s\approx M_{Planck}$}. As we said, this is the option which
is forced upon us in perturbative heterotic vacua \cite{rev}. In this case
$M_c\approx M_I$ and gauge coupling unification should take place also
about the same scale. There is a slight problem here since 
plane extrapolation of the MSSM gauge coplings indicate
unification at $2\times 10^{16}$ GeV, which is about a factor 20
too small compared to the string scale. This is naturally solved
identifying $M_s$ with the unification scale, which is option ii) :

{\bf ii) $M_s\approx M_X$} \cite{witten}. Here $M_X$ is the GUT scale
or the scale at which the extrapolated gauge couplings of the minimal  
supersymmetric standard model (MSSM) join. Numerically this is of order   
$10^{16}$ GeV. This corresponds (for the 3-brane case) to choices for
$M_c$ only slightly below $M_s$.

{\bf iii) $M_s\approx \sqrt{M_WM_{Planck}}$}. This is the geometrical   
intermediate scale $\approx 10^{11}$ GeV which coincides with the
SUSY-breaking scale in models with a hidden sector and gravity mediated
SUSY breaking in the observable sector. The interest of this choice has
been recently pointed out in ref.\cite{BIQ} (see also ref.\cite{benakli})
and recently explicit semirealistic D-brane models consistent
with this structure of mass scales have been constructed \cite{aiq} .
An interesting point of this possibility occurs in the case of
an isotropic compactification with $M_1=M_2=M_3=M_c$
with the SM embedded into 3-branes.
If there is a SUSY-breaking 3-brane sector
which is away in transverse space from the 3-branes 
containing the SM, the non-SUSY brane sector  acts as a standard hidden
SUSY-breaking  sector. In this case
one expects $M_{soft}=M_s^2/M_p=\alpha _3^2/2(M_c/M_s)^6$,
where $M_{soft}$ is the scale of SUSY-breaking felt
by the SM fields. Thus in order to have $M_{soft}\propto M_W$
 it is enough to have
$M_c/M_s\approx 0.01$, not very large relative scales are needed
\cite{BIQ} .
Thus the $M_W/M_p$ hierarchy may be naturally generated without 
any need for mechanisms like gaugino condensation.

{\bf iv) $M_s\approx 1$} TeV. This is the 1 TeV string scenario considered in
refs.\cite{lykken,untev,bacan}.
In this case it should be (in an isotropical 3-brane situation)
$M_c/M_s \approx 10^{-5}$. This case is potentially very exciting 
since the string scale could perhaps be testable at accelerator
energies. This  case has been discussed by other speakers
at this meeting \cite{bacan} .

We would like to argue that, within the context of D-brane models
the third and fourth options are phenomenologically safer \cite{imr} .
The reason is simple. A generic string theory background will 
contain in general D-brane systems with different amounts of supersymmetry.
Thus the SM model may perhaps be embedded
into a brane system which has $N=1$ supersymmetry. This could
guarantee that the standard gauge hierarchy problem is solved. 
However, generically there will be some other brane systems 
away in transverse directions which will have no supersymmetry ($N=0$).
If this is the case, the massless closed string states living in the
bulk of extra dimensions couple to both the $N=1$ brane sector
 (where the SM is contained) and the SUSY-breaking $N=0$ brane
sector.  Thus these closed string states will transmit supersymmetry
breaking to the SM sector supressed by the Planck mass and
of order $M_s^2/M_p$. If we want the magnitude of these soft
terms to be not bigger than the weak scale $M_W$ ( so that the 
hierarchy problem does not reappear) one needs to have:
\beq
 M_s \ \leq \ \sqrt{M_W M_p} \ \propto \ 10^{11} \ GeV
\label{prot}
\eeq
Thus, from the phenomenological point of view is safer to have 
$M_s\leq 10^{11}$ GeV in order to avoid too big SUSY-breaking effects from 
generic $N=0$ brane sectors. Of course, this is not a theorem, but is
quite suggestive. In fact, we will describe below a class of
orientifolds which provide a realization of this constraint.
Let me finally emphasize that in the cases iii) and iv) above
one will have to eventually find a mechanism to explain why
some of the compact dimensions became large compared to the
string size. It is this large size of some dimensions
which gives rise eventually to a $M_W/M_p$ hierarchy.

\section{ The gauge coupling unification problem }

If options iii) or iv) above are correct,
 the string scale would be well below
the standard grand unification scale $M_X=2\times 10^{16}$ GeV 
where the  unification of the  $SU(3)\times SU(2)\times U(1)$
couplings takes place when the minimal SUSY standard model spectrum 
is assumed to hold. Thus we have to face the following two 
important problems:

{\bf 1)} Couplings should unify at a scale $M_s\leq 10^{11}$ GeV. How we make
this compatible with the fact that the MSSM gauge couplings seem to
unify at a much larger scale of order $10^{16}$ GeV?

{\bf 2) }  Baryon and lepton number violating operators are supressed
only by inverse powers of $M_s$, which is now much lower than in the
conventional heterotic scenario. Thus unless apropriate symmetries
are present it is difficult to understand the level of proton stability
indicated by the experimental limits.

Concerning the first point, one has to emphasize that a detailed knowledge
of the gauge kinetic functions of the SM gauge groups is really
required in order to check whether coupling unification is still 
possible. Concerning the second point, one has to study whether 
apropriate symmetries could be present in order to supress sufficiently 
the operators violating proton stability 
(see ref.\cite{iq} ) .
For these two questions 
it turns out to be relevant the study of the pseudo-anomalous $U(1)$ 
gauge symmetries which are generically present in 
the class of Type IIB orientifold models mentioned
above. 

\subsection{Anomalous $U(1)$'s and mirage unification}

Indeed, if one computes the $U(1)$ triangle anomalies in Type IIB 
$D=4$ orientifolds one finds that most of the $U(1)$'s are anomalous.
This is not new in string theory: it is well known that in  heterotic 
string vacua there are analogous $U(1)$ symmetries whose triangle
anomalies are cancelled by a $D=4$ version of the Green-Schwarz 
mechanism \cite{dsw} 
. There is however a couple of important differences between the Type I
and heterotic cases. In the heterotic case there is only one anomalous
$U(1)$ and its mixed anomaly with all the non-Abelian gauge groups is identical.
This is because there is a single field (the complex dilaton $S$) giving 
rise to the GS mechanism. In addition a  Fayet-Iliopoulos (FI) 
term of order $g^2M_p^2/16\pi ^2$ appears at one-loop. The latter 
are of order the string scale. 
In the Type IIB orientifold models the story is quite different.
One finds that \cite{iru} 

i) There are multiple anomalous $U(1)$'s.

ii) The mixed anomalies of the $U(1)$'s with the different gauge
factors is non-universal.

iii) It is the twisted moduli fields $M_k$ which participate in the
GS mechanism, instead of the complex dilaton $S$.

iv) There appear FI-terms which are proportional to $<ReM_k>$, 
which are the "blowing-up" fields of the orbifold singularities. 
Thus, unlike the heterotic case, the FI terms may be arbitrarily small
\cite{iru,pop} .

 More specifically, cancellation of
$U(1)$ anomalies results \cite{iru} from the presence in the
$D=4$, $N=1$ effective action of the term
 \beq
\sum _k \delta^l_k B_k \wedge F_{U(1)_l} 
\eeq
where $k$ runs over the different twisted sectors of the
underlying orbifold (see ref.\cite{iru} for details) and
$B^k$ are the two-forms which are  dual to the imaginary
part of the twisted fields $M_k$. Here {\it l} labels the different
anomalous $U(1)$'s and $\delta^l_k$ are model-dependent constant
coefficients. In addition the gauge kinetic functions have
also a (tree-level) $M_k$-dependent piece:
\beq
f_{\alpha }\ = \ S  \ + \ \sum _k s^k_{\alpha } M_k
\label{correc}
\eeq
where the $s^k_{\alpha }$ are model dependent
coefficients. Under a $U(1)_l$ transformation the $M_k$
fields transform non-linearly
$
{\rm Im} M_k \ \rightarrow {\rm Im}  M_k \ +\ \delta^l_k\Lambda _l(x)
$.
 This non-linear transformation combined with eq.(\ref{correc})
results in the cancellation of the $U(1)$ anomalies
as long as the coefficients $C^l_{\alpha }$
 of the mixed $U(1)_l$-$G_{\alpha }^2$
anomalies are given by
\beq
C^l_{\alpha }\ =\ -\ \sum _k  s^k_{\alpha } \delta^l_k
\label{lares}
\eeq
Unlike the perturbative heterotic case, eq.(\ref{lares})
does not in general require universal mixed anomalies.

Now, 
equation \ref{correc}  shows us an interesting point
\cite{imr,mirage} : the gauge
coupling constants at the string scale in this class of theories
are only unified if one sits precisely at the orbifold 
points with $<ReM_k> =0$. Also, if $<ReM_k> \not= 0$
the corrections are group dependent and not universal. 
Thus consider a simplified scenario in which we we had
only a single blowing up field M 
so that $f_{\alpha } = S \ + \ s_{\alpha }M $. 
Consider now the renormalization group running of gauge couplings
$g_{\alpha}$ from the weak scale to the string scale $M_s$:
\beq
{{4 \pi}\over {g^2_{\alpha}(M_W) }}\ = \ {Re f_{\alpha}}
\ + \ { {  b_{\alpha } }\over {2\pi}} \log{ {M_s}\over {M_W} }
\label{running}
\eeq
where $f_{\alpha }$ is the gauge kinetic function in eq.(\ref{correc}).
We know that with the particle content of the MSSM coupling
unification works nicely for a unification scale $M_X=2\times
10^{16}$ GeV.
Thus if we had a model with:
\beq
 s_{\alpha } =  \gamma b_{\alpha }  \ ; \quad
\langle  Re M \rangle \ = \ {1 \over \gamma} {1 \over {2\pi}} \log(M_X/M_s)
\label{precoz}
\eeq
we would nicely get (aparent) gauge coupling unification.
This possibility may be named "mirage unification" 
\cite{mirage} because from
a low-energy observer, everything looks like if there was just
standard coupling unification at $M_X$
(for approaches similar in spirit see also 
refs.\cite{mormir} ) . In fact what happens is that
there are finite corrections to the gauge couplings 
at $M_s$ (which may be much smaller than $M_X$) which 
precisely mimic the effect. 
It turns out that there are indeed \cite{imr,mirage}  
orientifolds in which in 
some simple cases one can have $s_{\alpha }\propto \beta _{\alpha } $
(e.g., the $Z_3$ and $Z_7$ orientifolds). However in those cases,
the study of the scalar potential (including the FI terms) tells us
that $<M>=0$ at the minima with unbroken non-Abelian gauge group
\cite{mirage,abd}.
Thus in these orientifold examples the gauge couplings seem to unify at $M_X$,
mirage unification does not occurr. 
Nevertheless, odd orientifolds like these are very special and it could well
be that in more general situations the vacua may sit at points
with $<ReM>\not=0$. Notice also that once SUSY is broken
large non-vanishing D-terms will in general be allowed and 
$<ReM>$ may move from a vanishing value at the SUSY minimum to
a non-vanishing one after SUSY-breaking effects are taken into
account. 

\subsection{Precocious gauge coupling unification }

If mirage unification as above does not occurr and couplings join at
the string scale, one has to find an explanation for
"precocious" coupling unification at a scale $M_s<<10^{16}$ GeV.
The simplest and most conservative possibility is to abandon
the particle content of the MSSM and assume that there are
extra massless charged particles beyond quarks, leptons and one 
set of Higgs fields. If e.g. $M_s\propto  10^{10} - 10^{12}$ GeV, it is easy
to
find extra sets of particles which can have this effect. In particular,
a simple option is the addition of extra left-handed and right-handed
leptons which were shown in ref.\cite{BIQ}  to be consistent with
intermediate scale unification. This possibility   
could sound less natural than the MSSM paradigm with 
unification at $2\times 10^{16}$ GeV but one should be
more open minded and not  look at that paradigm as the unique possibility. 
Let us remind that the MSSM structure is in some respects a bit artificial:
whereas quarks and leptons come in three chiral copies, Higgsses come
in only one (vector-like) copy. Furthermore, symmetries 
have to be
impossed in order to insure  sufficient proton stability.
 Interestingly enough it has been recently been found 
\cite{aiq,aiq2} that in 
semi-realistic Type -I models there is a tendency to get extra massless
leptons which could lead to gauge coupling unification at an
intermediate scale (see below).

\section{Standard-like models from Type I string vacua}

It turns out to be quite difficult to construct semi-realistic
$D=4$ compact orientifolds with unbroken $N=1$ supersymmetry.
As we mentioned above, the tadpole cancellation constraints 
are so strong that there is little flexibility left for
obtaining a realistic gauge group and three quark-lepton
generations. One obvious direction in order to obtain more 
flexibility is to consider also non-supersymmetric vacua
: after all the world is not (exactly) supersymmetric. 
This possibility was excluded in the past because of the hierarchy problem:
if SUSY is broken at the string scale, with the latter close to the
Planck scale, no scalar would survive radiative corrections
and we would be left with no Higgs fields in order to
break the $SU(2)_L\times U(1)$ symmetry.

Since we can now lower the string scale well below the
Planck mass, the above exclusion of non-supersymmetric vacua
must be reexamined. There are now two new possibilities
opened: 1) Having a non-SUSY model with the string scale
not much above the weak scale or 2) Having a non-SUSY model
with the string scale of order the intermediate scale. In the latter case if
the non-SUSY sector of the theory is only connected to the
SM world by the exchange of bulk (closed string) fields, the
hierachy can in general be preserved.

Recently \cite{aiq,aiq2}  Type IIB , $D=4$ 
compact orientifolds providing for  explicit realizations of the
above possibilities have been obtained for the first time.
 They are based on the observation \cite{ads,au} that 
one can obtain tadpole- and tachyon-free configurations by adding 
brane-antibrane pairs to $N=1$, $D=4$ orientifolds. The presence of
the anti-branes makes  this kind of configuration 
 non-supersymmetric.
Let us present an specific orientifold example \cite{aiq}  
based on the construction in ref.\cite{au} yielding a
semirealistic spectrum. It is based in the standard $Z_3$
orientifold constructed with 7-branes instead of
9-branes.  One compactifies the Type IIB string on the standard
$Z_3$ orbifold. The orientifold projector is given by
$\Omega (-1)^{F_L}R_3$, where $R_3$ is the operation reflecting the third
compact complex plane. Tadpole cancellation conditions require the presence
of 32 7-branes with their worldvolume including the first two complex planes
plus Minkowski space
. Now, we embedd the $Z_3$ action into the 
7-brane Chan-Paton factors by chosing \cite{aiq} :
\beq
V_7 \ = \
1/3(1,1,1,-1,-1,0,0,0,0,1,1,1,1,1,1,1)
 \eeq 
In addition a quantized Wilson
line is added in the first complex plane given by:
\beq
W_7 \ = \
1/3(1,1,1,1,1,1,1,0,0,0,0,0,0,0,0,0)
 \eeq 
The gauge group from the
(77) sector will be $SU(3)\times SU(2)_L\times S U(2)_R \times
U(1)_{B-L} 
\times [U(1)^2\times SO(4)\times U(7)]$.
This contains the left-right symmetric extenssion of the
SM which is a phenomenologically interesting model.
One can also check that from the
(77) sector there are chiral fields transforming like:
\beq
3(3,2,1,1/3)\ +\ 3({\bar 3},1,2,-1/3) \ +\ 3(1,2,2,0)  
\eeq
under $SU(3)\times SU(2)_L\times S U(2)_R \times
U(1)_{B-L}$ (plus extra fields transforming under the
hidden group)
. These are three quark generations plus three 
sets of Higgs fields. 
Now, the nine fixed points under $Z_3$ in the
first two complex planes split into  three sets
of three fixed points each  which have associated twists $V$,
$V+W$ and $V-W$ respectively. 
The corresponding value for $\Tr
\gamma_{\theta,7}$ are -4, -4 and -1. 
Now, for this orientifold tadpole cancellation conditions require
 $\Tr \gamma_{\theta,7}=-4$. This means that we will have to add
something on the three fixed points with shift $V-W$ in order to 
cancel tadpoles. 
It is easy to see that if we locate
at each point  two  3-branes with
   $\Tr \gamma_{\theta,3}= -1$ all twisted tadpoles cancel.
Now there are extra massless chiral fields from open strings 
extending between the 3-branes in the three fixed points and the 7-branes.
They will have gauge quantum numbers under the (77) gauge group:
\beq
(1,2,1,+1) \ +\
(1,1,2,-1)\ +\ (3,1,1,-2/3)+({\bar 3},1,1,2/3) \ 
 \eeq 
plus extra fields transforming under the hidden group.
These come in three copies (one per fixed point). 
Notice that in this sector three standard lepton generations
appear. In addition there are three extra sets of vector-like 
colour triplets which turn out to get generically large 
masses (see \cite{aiq,aiq2}  for details).
Thus we have easily constructed a three generation 
left-right symmetric model starting with the simplest
$Z_3$ orientifold and adding apropriate numbers of
7-branes and 3-branes. This is, by the way, the simplest
semi-realist string model I have ever seen.
Notice that this model has three generations of
Higgs fields. This is a general trend in these constructions, 
there are typically extra massless weakly interacting
fields. This is interesting because they lead to
precocious gauge coupling unification, $SU(2)_L$ and 
$U(1)_Y$ interactions grow faster than in the MSSM 
and tend to join at an intermediate scale
$10^{8}-10^{12}$ GeV with the $SU(3)$ coupling
\cite{aiq,aiq2} .
Thus this class of models provide a natural
alternative to the MSSM scenario in which
couplings join close to the Planck mass.

This model is non-supersymmetric because there is an
additional tadpole cancelation condition: the net-number of
3-branes minus anti-3-branes must be zero in this model.
Thus there must be 6 anti-3-branes somewhere. Depending on 
the location of these extra anti-3-branes, the SUSY-breaking 
phenomenology is different. Let us locate for definiteness
the 7-branes at the origin in the third complex dimension. 
Now, if the anti-3-branes are away from the origin in the
third compact dimension, they have no overlap with the 7-branes 
and hence there are no massless chiral fields in
the $({\bar 3}7)$ sector. In this case the SUSY-breaking 
spectrum residing in this anti-3-branes can only 
communicate with the (77) and (37) sectors (which
contain the observed physics) by the exchange of
closed string fields which live in the bulk.
The couplings of the latter are supressed by
powers of the Planck mass. In this case the model
behaves like the standard hidden sector SUSY-breaking models
in which the role of hidden sector is played by the anti-3-branes.
For this to work the string scale must be the intermediate scale.
This can be made consistent with the observed Planck mass
by e.g., chosing compactification scales $M_i$ along the three
complex compact dimensions as follows: $M_1\propto M_2=\propto M_s \propto
10^{11}$ GeV and $M_3\propto 1$ TeV.
Alternatively, if the anti-3-branes are located at the origin 
in the third complex plane, their worldvolume will overlap 
with that  of 7-branes and there will be a non-SUSY massless
spectrum coupling to the SM gauge group. In this case, if we do not want to
have a hierarchy problem, one should lower the string scale 
down to  $M_s\propto 1-10$ TeV. This is again possible by chosing
$M_1\propto M_2\propto M_s\propto 1-10 $ TeV and $M_3\propto 
10^{-3}$ eV. 

Since this class of models, although free of Ramond-Ramond tadpoles
and tachyons, are non-supersymmetric, their stability should
be farther studied. Notice however that this is a problem that
we will have to face anyhow in any semirealistic model. In the 
traditional heterotic models one had to resort to field theory
effects like gaugino condensation to break supersymmetry
in a hidden sector, and this leads to the same questions that
we face now in the non-supersymmetric Type IIB orientifolds.
One of the adventages now is that those effects are produced
by explicit anti-D-branes whose effects can in principle
be better studied.

\section{Outlook}

We are witnessing at the moment something we could perhaps name
(to follow the tradition) the second string (phenomenology) revolution.  
The heterotic string has lost its monopoly as the  candiate
 for the unification of gravity and the standard model of particle
physics. Although M-theory is supossed to be the unique underlying
theory, one of the different perturbative limits like Type I, Type II 
and heterotics could perhaps be closer than the others to the
observed physics. In the last fifteen years essentially only the
heterotic string has been explored, with 
important (but limited) success. The first
Type I semirealistic models are now starting to be built and
they show some very interesting features compared to their
heterotic precursors. One of them is the possibility that
the string scale is much below the Planck mass, at the intermediate
scale $\sqrt{M_WM_p}$ or even close to the weak scale.
These are first steps in the search for realistic 
string vacua using D-brane techniques. Much  work 
remains to be done both from the
theoretical and more phenomenological sides.

\bigskip

\section*{Acknowledgments}

I am grateful to  G. Aldazabal, 
A. Font , C. Mu\~noz, F. Quevedo,
R. Rabadan, S. Rigolin,  A. Uranga and G. Violero for an enjoyable
collaboration.


\end{document}